\documentclass[useAMS,usenatbib,referee]{biom2arxiv}

\usepackage{tikz}
\usetikzlibrary{arrows}
\usepackage{bbm}
\usepackage{bm}
\usepackage{amsmath, mathtools}
\usepackage{amssymb}
\usepackage{graphicx}
\usepackage{enumitem}
\setlist{nolistsep,noitemsep,topsep=0cm,after=\vspace{0cm},before=\vspace{0cm}}
\usepackage{xr}
\usepackage{color}
\usepackage{url}

\usepackage{array,multirow}

\newcommand{\beq}{\begin{equation}}
\newcommand{\eeq}{\end{equation}}
\newcommand{\ben}{\begin{enumerate}}
\newcommand{\een}{\end{enumerate}}
\newcommand{\bei}{\begin{itemize}}
\newcommand{\eei}{\end{itemize}}

\newcommand{\indicator}[1]{{\mathbbm 1}\left(#1\right)}
\newcommand\ind{\protect\mathpalette{\protect\independenT}{\perp}}
\def\independenT#1#2{\mathrel{\rlap{$#1#2$}\mkern2mu{#1#2}}}

\newcommand{\PreserveBackslash}[1]{\let\temp=\\#1\let\\=\temp}
\newcolumntype{C}[1]{>{\PreserveBackslash\centering}p{#1}}

\DeclareMathOperator*{\E}{E}

\title{Non-linear Mediation Analysis with High-dimensional Mediators whose Causal Structure is Unknown}

\author{
Wen Wei Loh$^{1,*}$\email{WenWei.Loh@UGent.be},
Beatrijs Moerkerke$^{1}$, 
Tom Loeys$^{1}$, and 
Stijn Vansteelandt$^{2,3}$ \\
$^{1}$ Department of Data Analysis, Ghent University, Gent, Belgium\\
$^{2}$ Department of Applied Mathematics, Computer Science and Statistics, Ghent University, Ghent, Belgium \\
$^{3}$ Department of Medical Statistics, London School of Hygiene and Tropical Medicine, United Kingdom \\
}

\begin{document}

\label{firstpage}

\begin{abstract}

With multiple possible mediators on the causal pathway from a treatment to an outcome, we consider the problem of decomposing the effects along multiple possible causal path(s) through each distinct mediator. 
Under Pearl's path-specific effects framework \citep{pearl2001direct,avin2005identifiability}, such fine-grained decompositions necessitate stringent assumptions, such as correctly specifying the causal structure among the mediators, and no unobserved confounding among the mediators. 
In contrast, interventional direct and indirect effects for multiple mediators \citep{vansteelandt2017interventional} can be identified under much weaker conditions, while providing scientifically relevant causal interpretations.
Nonetheless, current estimation approaches require (correctly) specifying a model for the joint mediator distribution, which can be difficult when there is a high-dimensional set of possibly continuous and non-continuous mediators.
In this article, we avoid the need to model this distribution, by developing a definition of interventional effects previously suggested by \citet{vanderweele2017mediation} for longitudinal mediation. 
We propose a novel estimation strategy that uses non-parametric estimates of the (counterfactual) mediator distributions.
Non-continuous outcomes can be accommodated using non-linear outcome models. Estimation proceeds via Monte Carlo integration.
The procedure is illustrated using publicly available genomic data \citep{huang2016hypothesis} to assess the causal effect of a microRNA expression on the three-month mortality of brain cancer patients that is potentially mediated by expression values of multiple genes.

\end{abstract}

\begin{keywords}
Collapsibility;
Direct and indirect effects; 
Effect decomposition;
Marginal and conditional effects;
Multiple mediation analysis;
Path analysis
\end{keywords}

\maketitle

\section{Introduction}
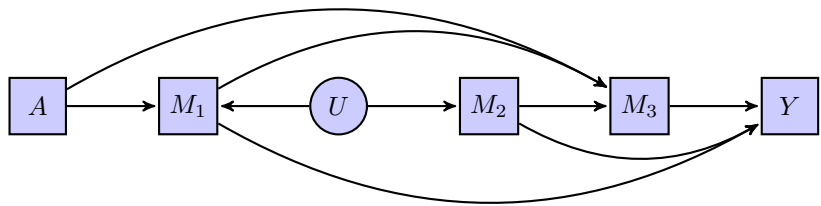
\begin{figure}
\centering
\caption{Causal diagram for an example with three mediators, two of which share an unobserved confounder $U$. 
Observed variables are drawn as rectangular nodes and unobserved variables are drawn as round nodes.
\label{fig-hdexample}}
\begin{tikzpicture}[->,>=stealth',shorten >=1pt,auto,node distance=2cm,thick,
  latent node/.style={circle,fill=blue!20,draw,
  font=\sffamily\large\bfseries,minimum size=.75cm},
  main node/.style={rectangle,fill=blue!20,draw,
  font=\sffamily\large\bfseries,minimum size=.75cm}]

  \node[main node] (A) {$A$};
  \node[main node] (M1) [right of=A] {$M_1$};
  \node[latent node] (U) [right of=M1] {$U$};  
  \node[main node] (M2) [right of=U] {$M_2$};
  \node[main node] (M3) [right of=M2] {$M_3$};
  \node[main node] (Y) [right of=M3] {$Y$};
  
  \path[black]
  	(A) edge (M1)
  	(A) edge[bend left] (M3)	
  	(M1) edge[bend right] (Y)
  	(M2) edge (M3)		
  	(M3) edge (Y)			
  	(U) edge (M1)
  	(U) edge (M2)		
  	(M2) edge[bend right] (Y)	
  	(M1) edge[bend left] (M3)	
	;	
\end{tikzpicture}
\end{figure}

Mediation analysis is commonly used to study the effect of a treatment or exposure ($A$) on an outcome ($Y$) that may be transmitted through intermediate variable(s) on the causal pathway from $A$ to $Y$. 
Counterfactual-based definitions of {\em natural direct and indirect effects} \citep{robins1992identifiability,pearl2001direct} permit decomposing the total effect of a treatment on an outcome into a direct and an indirect effect for a single mediator, without relying on any specific statistical models for the mediator and outcome. 
However, multiple possible mediators often exist in substantive research.
For example, different mediators may be posited in trying to understand the different causal pathways from $A$ to $Y$, or confounders of the mediator-outcome relation for a mediator of interest are themselves affected by treatment and thus perceived as competing mediators, or interventions may be designed to affect outcome by simultaneously changing different mediators on the causal pathway from $A$ to $Y$.
Extensions of natural effects for a single mediator to the multiple mediator setting are therefore complicated by the complex (possibly unknown) confounding patterns among the different mediators.
In particular, when one mediator exerts a causal effect on another, so that the former is a confounder of the mediator-outcome relation for the latter (henceforth termed {\em post-treatment} confounding), the assumptions needed to (non-parametrically) identify the natural or separate {\em path-specific} effects through each mediator are violated \citep{avin2005identifiability}. 
Consider the following example in Figure~\ref{fig-hdexample}. 
The natural indirect effect via $M_3$ alone cannot be identified because $M_1$ and $M_2$ are post-treatment confounders of the $M_3-Y$ relation.
The path-specific indirect effect along the assumed path $A \rightarrow M_1 \rightarrow M_2 \rightarrow Y$ cannot be identified because $U$ is an unobserved confounder of $M_1$ and $M_2$.

Current methods for assessing natural effects with multiple mediators are restricted to situations where either the mediators can be causally ordered so that (combinations of) path-specific effects can be identified in the presence of post-treatment confounding \citep{vanderweele2014mediation,daniel2015causal,steen2017flexible,albert2019gmediation}, or the mediators do not exert causal effects on each other, and are independent given treatment and covariates \citep{lange2013assessing,taguri2018}.
Hence a limitation, shared by all the above approaches, is that they are predicated on (correct) a priori assumptions about the detailed causal relations between the mediators.
These restrictions impede applications in most realistic settings, when the causal structure among the mediators is unknown or cannot be (correctly) specified based on sound scientific knowledge or empirical evidence, or the mediators are manifestations of a latent process or variable, and thus correlated. 

In contrast, {\em interventional direct and indirect effects}, first introduced by \citet{didelez2006direct} and \citet{vanderweele2014effect} for a single mediator, can be identified under much weaker assumptions than natural or path-specific effects, especially when there is post-treatment confounding of the mediator(s)-outcome relation(s).
Interventional effects consider population-level (stochastic) interventions that set the value of the mediator to a random draw from its (counterfactual) distribution; whereas natural effects are defined in terms of individual-level (deterministic) interventions on the mediator, which can lack scientific and practical meaning when the treatment cannot be manipulated at the individual level \citep{vanderweele2013policy}.
\citet{vansteelandt2017interventional} extended the definitions to the multiple mediator setting so that the total effect of a treatment on an outcome can be decomposed into a direct effect and a joint indirect effect via the mediators. They further decomposed the joint indirect effect into separate indirect effects via each distinct mediator, and an indirect effect via the mediators' mutual dependence. 
Interventional indirect effects are defined in terms of the underlying (possibly unknown) causal effects among the mediators. They therefore possess valid interpretations by construction, and can be identified without prior (correct) specification of the mediators' causal structure, and even when mediators share hidden common causes. 
Recent work discussing interventional (in)direct effects include \citet{moreno2018understanding}, \citet{lok2019organic}, and \citet{trang2019clarifying}, among others, for a single mediator; \citet{vanderweele2017mediation} for a single longitudinal mediator;
\citet{lin2017interventional} for multiple mediators with a known causal ordering; and \citet{loh2019disentangling} for comparisons with prevalent ``product-of-coefficients'' methods \citep{mackinnon2000contrasts,preacher2008asymptotic} assuming linear models for all variables.

In this article, we generalize the interventional effects framework for multiple mediators to high-dimensional mediators. A key complication when estimating current definitions of the interventional effects is the need to correctly specify a (parametric) model for the joint distribution for all mediators conditional on treatment and all observed confounders.
However, \citet{vanderweele2017mediation} suggested the possibility of interventional effects defined using (counterfactual) mediator distributions that depend only on treatment, and are unconditional on the baseline confounders of the mediator(s)-outcome relation(s).
Such definitions avoid specifying (marginal structural) models for the mediators in terms of the baseline confounders, and can thus be particularly advantageous when there is a high-dimensional set of (non-)continuous mediators.
All current high-dimensional mediation methods for assessing indirect effects through each distinct mediator are restricted to continuous mediators, with almost all additionally considering only continuous outcomes, so that the product-of-coefficients method under assumed linear models for (transformations of the) mediators and outcomes can be employed;
see e.g., 
\citet{chakrabortty2018inference},
\citet{huang2016hypothesis},
\citet{zhang2016estimating},
\citet{zhao2016pathway},
\citet{zhao2019multimodal},
\citet{derkach2020group}, and
\citet{zhao2020sparse}
among many others.
While most methods allow for either correlated errors in the joint mediator model (thus allowing for certain forms of hidden confounding among the mediators), or mediators to influence one another, only \citet{zhao2016pathway} and \citet{zhao2019multimodal} allow for both within the same mediation model. 
In most realistic and important settings, high-dimensional mediators that are closely linked either have complex but unknown causal relations, or share unmeasured common causes, or both. 
Even in the single mediator setting, indirect effects using the traditional product-of-coefficients method may not correspond to those derived using the causal mediation framework, due to misspecification of non-linear models for non-continuous mediator and outcome \citep{mackinnon2018correspondence}.
With the increasing demand for high-dimensional mediation methods in biology, medical and public health research, our proposal is, to the best of our knowledge, the first to allow for both continuous and non-continuous mediators to simultaneously exist on the causal pathway between treatment and outcome, where they can concurrently causally affect one another, and share hidden common causes.

In this article, we therefore build on the suggestion by \citet{vanderweele2017mediation} for a single (longitudinal) mediator to develop interventional direct and indirect effects for high-dimensional mediators.
Our proposed definitions of interventional indirect effects differ from existing definitions by \citet{vansteelandt2017interventional}, \citet{lin2017interventional}, and \citet{loh2019disentangling} in two respects. The first is that the decomposition of the joint indirect effect into separate indirect effects via each mediator is invariant to the presumed (possibly arbitrary) ordering of the mediator indices. Existing definitions assume different hypothetical treatment levels for different subsets of the mediators, thus allowing the possibility of different decompositions (depending on the mediator indices).
The second and more pertinent difference is that the interventional indirect effects developed in this article rely on (counterfactual) mediator distributions that depend only on treatment, whereas existing definitions depend on treatment and all baseline covariates. 
In high-dimensional mediation settings, specifying a (correct) model for the joint distribution for all mediators that is congenial with models for the marginal distribution for each mediator, all of which are conditional on the covariates, can be difficult.
The definitions proposed in this article therefore permit a novel estimation strategy that requires specifying only a (marginal structural) mean model for the outcome, and no models for the mediators. 
The mediators and outcome can be continuous or noncontinuous, with non-continuous outcomes accommodated using non-linear outcome models. Non-parametric estimates of the (counterfactual) mediator distributions under each treatment level are used.
Estimation proceeds via Monte Carlo integration.

The remainder of this article is as follows. 
In Section~\ref{sect:notation_definition} notation is introduced, the interventional direct and indirect effects are defined, and the identification assumptions are stated.
In Section~\ref{sect:effectmodels} the novel estimation strategy that requires no models for the mediators, and only a mean model for the outcome, is proposed. 
For pedagogic purposes, we will focus on randomized studies with fewer mediators than observations in the theoretical development of our proposed procedure. Extensions to high-dimensional mediators, and observational studies when treatment is not randomly assigned, are then described.
In Section~\ref{sect:simstudies} the proposed estimation procedures are assessed via extensive simulation studies.
In Section~\ref{sect:application} the estimation strategy is illustrated using publicly-available data from a previous high-dimensional mediation analysis \citep{huang2016hypothesis} that investigated the causal effect of a microRNA expression (miR-223) on a dichotomous three-month survival status among patients suffering from an aggressive form of brain cancer, that is potentially mediated by expression values of different genes.
In Section~\ref{sect:compare} we demonstrate that in most common settings, the proposed estimation procedure simplifies current estimation approaches \citep{vansteelandt2017interventional,loh2019disentangling}, by eliminating the need to specify a model for the joint distribution of the mediators. A brief discussion is provided in Section~\ref{sect:discussion}.

\section{Definition and identification of interventional (in)direct effects \label{sect:notation_definition}}

Consider the setting with an exposure or a treatment $A$, multiple possible mediators $M_1, \ldots, M_p$, and an outcome $Y$. In this article, we adopt the perspective that all post-treatment confounders of a mediator-outcome relation for a mediator in question are themselves competing mediators, and therefore included in the set of possible mediators. Here and throughout subscripts in the notation for the mediators are merely used to arbitrarily index the different mediators, and not to indicate any assumed causal (or temporal) ordering of the mediators; e.g., $M_1$ need not precede $M_2$ causally (or temporally).
Let $Y_{a m_1\cdots m_p}$ denote the potential outcome for $Y$ if, possibly counter to fact, $A$ is set to $a$, when each mediator $M_s$ is set to the value $m_s, s=1,\ldots,p$. Let $M_{sa^{(s)}}$ denote the potential outcome for $M_s$ if, possibly counter to fact, $A$ is set to $a^{(s)}$.
Let $Y_{a^{(0)}\{\tilde M_{1a^{(1)}}\cdots\tilde M_{pa^{(1)}}\}}$ denote the potential outcome for $Y$ under treatment $A=a^{(0)}$, when the mediator values are set to a random draw from the {\em joint} (counterfactual) distribution under (hypothetical) treatment $A=a^{(1)}$, i.e., $\{\tilde M_{1a^{(1)}} \cdots \tilde M_{pa^{(1)}}\} \sim F(M_{1a^{(1)}},\ldots,M_{pa^{(1)}})$, where $F(X)$ denotes a cumulative distribution function for $X$.
Potential outcomes where the mediators are set to random draws from the joint counterfactual distribution are henceforth denoted by (curly) brackets in the subscripts. 
Let $Y_{a^{(0)}\tilde M_{1a^{(1)}}\cdots\tilde M_{pa^{(p)}}}$ denote the potential outcome under treatment $A=a^{(0)}$, when the value of each mediator is set to a random draw from the {\em marginal} (counterfactual) distribution that does not depend on any other mediator, i.e., $\tilde M_{sa^{(s)}} \sim F(M_{sa^{(s)}}), s=1,\ldots,p$.
Let $L$ denote all observed baseline (i.e., unaffected by treatment) covariates that may affect any of $(A,M_1,\ldots,M_p, Y)$.
Here and throughout, the joint and all marginal distributions for the (counterfactual) mediators are unconditional on $L$.
The average potential outcomes (hereafter termed ``estimands'') are respectively defined as:
\begin{align}
\E\left(Y_{a^{(0)}\{\tilde M_{1a^{(1)}}\cdots\tilde M_{pa^{(1)}}\}}\right)
&=
\int \E(Y_{a^{(0)} m_1\cdots m_p}) \; 
dF_{M_{1a^{(1)}},\cdots, M_{pa^{(1)}}}(m_1, \cdots, m_p), \quad \mbox{and}
\label{eq:estimand_joint_unconditional}\\
\E\left(Y_{a^{(0)}\tilde M_{1a^{(1)}}\cdots\tilde M_{pa^{(p)}}}\right)
&=
\int \E(Y_{a^{(0)} m_1\cdots m_p}) \;
dF_{M_{1a^{(1)}}}(m_1)\cdots dF_{M_{pa^{(p)}}}(m_p).
\label{eq:estimand_marginal_unconditional}
\end{align}

\subsection{Interventional (in)direct effects for multiple mediators \label{sect:effects_definition}}

The interventional effects comparing estimands \eqref{eq:estimand_joint_unconditional} and \eqref{eq:estimand_marginal_unconditional} under different hypothetical treatment levels for a binary treatment $A$ are defined as follows. Let $g$ denote a user-specified link function, such as 
the log link $g(x)=\log(x)$, or the ``logit'' link $g(x)=\log\{x/(1-x)\}$. 
Define the total effect as:
\[
g\left\{\E\left(Y_{1\{\tilde M_{11}\cdots\tilde M_{p1}\}}\right)\right\} - 
g\left\{\E\left(Y_{0\{\tilde M_{10}\cdots\tilde M_{p0}\}}\right)\right\},
\] 
which can be decomposed into the direct effect, defined as:
\beq\label{eq:DE_define_marginal}
g\left\{\E\left(Y_{1\{\tilde M_{11}\cdots\tilde M_{p1}\}}\right)\right\} - 
g\left\{\E\left(Y_{0\{\tilde M_{11}\cdots\tilde M_{p1}\}}\right)\right\},
\eeq
and the (joint) indirect effect, defined as:
\beq\label{eq:IE_define_marginal}
g\left\{\E\left(Y_{0\{\tilde M_{11}\cdots\tilde M_{p1}\}}\right)\right\} - 
g\left\{\E\left(Y_{0\{\tilde M_{10}\cdots\tilde M_{p0}\}}\right)\right\}.
\eeq
The indirect effect via each mediator $M_s, s=1,\ldots,p$, is defined as: 
\beq\label{eq:IEs_define_marginal}
g\left\{\E\left(Y_{0\tilde M_{10}\cdots\tilde M_{s-1,0} \, \tilde M_{s1} \, \tilde M_{s+1,0}\cdots\tilde M_{p0}}\right)\right\} - 
g\left\{\E\left(Y_{0\tilde M_{10}\cdots\tilde M_{p0}}\right)\right\}.
\eeq
Changing only the {\em marginal} counterfactual distribution of $M_s$ from treatment to control implies an ``overall'' effect of treatment on $M_s$ that is unconditional on any other mediators.
This effect therefore marginalizes over all (underlying) causal effects along (unknown) paths from treatment to the mediator in question that may intersect other causally antecedent mediators.
Hence the interventional indirect effect through a mediator can be readily interpreted as the combined causal effect along all (underlying) paths from treatment to the mediator in question, then directly to the outcome. 
Continuing the example shown in Figure~\ref{fig-hdexample}, the indirect effect through $M_3$ is the combined effect along the paths $A \rightarrow M_3 \rightarrow Y$ and $A \rightarrow M_1 \rightarrow M_3 \rightarrow Y$.
Further interpretations are deferred to the simulation studies in Section~\ref{sect:simstudies} and the illustration in Section~\ref{sect:application}.

Note that in the first term of \eqref{eq:IEs_define_marginal}, only the mediator in question $M_s$ is drawn from its distribution under treatment $a^{(s)}=1$; all other mediators are drawn from their distributions under control, i.e., $a^{(k)}=0$ for $k=1,\ldots,p, k \neq s$.
This definition of the indirect effect \eqref{eq:IEs_define_marginal} is especially amenable to settings with high-dimensional mediators because it does not require fixing the hypothetical treatments for the other mediators at different levels, and is hence invariant to the chosen (possibly arbitrary) ordering of the mediator indices.
In contrast, existing definitions \citep{vansteelandt2017interventional,loh2019disentangling} set $a^{(k)}=1$ for $k=1,\ldots,s-1$, and $a^{(k)}=0$ for $k=s+1,\ldots,p$, so that different definitions are possible depending on the mediator indices, although the conceptual interpretations remain the same.

Lastly, the difference between the joint indirect effect \eqref{eq:IE_define_marginal} and the sum of the separate indirect effects \eqref{eq:IEs_define_marginal} for all mediators can be further partitioned into:
\begin{align}
&\left[g\left\{\E\left(Y_{0\{\tilde M_{11}\cdots\tilde M_{p1}\}}\right)\right\}
-g\left\{\E\left(Y_{0\tilde M_{11}\cdots\tilde M_{p1}}\right)\right\}\right] \notag\\
&\quad-
\left[g\left\{\E\left(Y_{0\{\tilde M_{10}\cdots\tilde M_{p0}\}}\right)\right\}
-g\left\{\E\left(Y_{0\tilde M_{10}\cdots\tilde M_{p0}}\right)\right\}\right],
\label{eq:IEmutual_define_marginal} \\[12pt]
\mbox{and} \quad\quad
&\left[g\left\{\E\left(Y_{0\tilde M_{11}\cdots\tilde M_{p1}}\right)\right\}
- g\left\{\E\left(Y_{0\tilde M_{10}\cdots\tilde M_{p0}}\right)\right\} \right] \notag\\
&\quad-
\sum_{s=1}^{p}
\left[g\left\{\E\left(Y_{0\tilde M_{10}\cdots\tilde M_{s-1,0} \, \tilde M_{s1} \, \tilde M_{s+1,0}\cdots\tilde M_{p0}}\right)\right\}
-g\left\{\E\left(Y_{0\tilde M_{10}\cdots\tilde M_{p0}}\right)\right\}\right].
\label{eq:IEnegsum_define_marginal}
\end{align}
Following \citet{vansteelandt2017interventional}, we refer to \eqref{eq:IEmutual_define_marginal} as the indirect effect via the mediators' {\em mutual dependence}.
This indirect effect describes how treatment affects the relationships between the mediators, which subsequently affects the outcome, and thus cannot be attributed to any single mediator. 
We term the indirect effect in \eqref{eq:IEnegsum_define_marginal} as the ``remainder'' effect after removing the indirect effect via the mutual dependence \eqref{eq:IEmutual_define_marginal} from the difference between the joint indirect effect \eqref{eq:IE_define_marginal} and the sum of the separate indirect effects \eqref{eq:IEs_define_marginal}. 
Interpretations and closed-form expressions for the indirect effects \eqref{eq:IEmutual_define_marginal} and \eqref{eq:IEnegsum_define_marginal}, under the setting with two mediators and (non-)linear models for the mediators (and outcome), are provided in the Web Appendix.
For example, when the mean model for the outcome is linear and the mediators are normally distributed, the indirect effect via the mediators' mutual dependence \eqref{eq:IEmutual_define_marginal} is non-zero if and only if the covariance of the mediators is affected by treatment and the mediator-mediator interaction effect on the outcome is non-zero. 
When the mean model for the outcome is log-linear and the mediators are normally distributed, this indirect effect is non-zero if and only if the covariance of the mediators is affected by treatment and the main effects of the mediators on the outcome are both non-zero.

\subsection{Assumptions for identification \label{sect:identification}}
Identification of the interventional effects requires the following assumptions, where ``$\ind$'' denotes conditional independence:
\begin{align}
Y_{a m_1\cdots m_p} &\ind A | L \quad \forall \; a, m_1, \ldots, m_p; \label{eq:identify_1}\\
Y_{a m_1\cdots m_p} &\ind \{M_1 \cdots M_p\} | (A=a, L)  \quad \forall \; a, m_1, \ldots, m_p; \label{eq:identify_2}\\
\{M_{1a} \cdots M_{pa}\} &\ind A | L \quad \forall \; a. \label{eq:identify_3}
\end{align}
Assumptions \eqref{eq:identify_1} and \eqref{eq:identify_3} state that the effect of treatment $A$ on outcome $Y$, and the effects of treatment $A$ on all mediators,  are unconfounded conditional on $L$.
Assumption~\eqref{eq:identify_2} states that there is sufficient information observed in $L$ so that the association between any of the mediators $(M_1,\ldots,M_p)$ and outcome $Y$ is unconfounded within levels of the covariates $L$. 
Because this assumption is not empirically testable, all baseline measurements of the mediators and the outcome (prior to treatment being received) should be adjusted for in practice, even when treatment is randomly assigned.
We refer readers to \citet{smith2019mediational} for implications when this assumption is violated, and recommended sensitivity analyses.

\vspace{-.2in}

\section{Estimation of interventional (in)direct effects \label{sect:effectmodels}}
In this section we develop a novel estimation strategy that requires only an outcome model, and no models for the mediators. The strategy exploits the proposed definitions of the interventional effects using (counterfactual) mediator distributions that are unconditional on any covariates. 
For pedagogic purposes, we will first assume that there are sufficient observations for an unpenalized outcome model, conditional on treatment, all mediators, and covariates, to be fitted to the observed data. 
We will further assume that treatment is randomly assigned so that both independence assumptions \eqref{eq:identify_1} and \eqref{eq:identify_3} are satisfied unconditionally on $L$. Extensions to high-dimensional mediators, and observational studies with non-randomly assigned treatments, are presented in later sections.

\subsection{Randomly assigned treatment with an unpenalized outcome model \label{sect:lowdimensional_random}}

Estimators of the proposed interventional (in)direct effects are obtained as follows:

\ben[label=A\arabic*.]\setcounter{enumi}{-1}
\item Fit an outcome model, conditional on treatment, mediators, and covariates, to the observed data, e.g., $\E(Y|A,M_1,\ldots,M_p,L)$. The outcome model can be expressed as a function of its inputs, e.g., $\E(Y|A=a,M_1=m_1,\ldots,M_p=m_p,L)=h(a,m_1, \ldots, m_p, L)$, where $h(\cdot)$ is a user-specified function. 
The observed values of the covariates $L$ for each individual are assumed to be fixed and invariant to different values of the treatment $a$ and counterfactual mediator values (denoted by $m_1, \ldots, m_p$). Denote the estimated function by $\hat h(a,m_1, \ldots, m_p, L)$.
\item Construct the duplicated data for each individual as shown in Table~\ref{table:duplicatedall_fitM1}.
The hypothetical treatment levels $a^{(0)}$ and $a^{(1)}$ are chosen so that the interventional direct effect \eqref{eq:DE_define_marginal} is the difference between the (transformed) estimands in the last and penultimate rows, and the joint indirect effect \eqref{eq:IE_define_marginal} is the difference between the (transformed) estimands in the penultimate and first rows.

\begin{table}
\caption{Duplicated data for each individual when estimating the direct and joint indirect effects. The counterfactual mediators $\{\tilde M_{1a^{(1)}} \cdots \tilde M_{pa^{(1)}}\}$ are randomly drawn (jointly) from the observed treatment group $A=a^{(1)}$. The column $L$ is omitted for simplicity. \label{table:duplicatedall_fitM1}}
\begin{center}
\begin{tabular}{ |c|c|ccc|c| } 
 \hline
 $a^{(0)}$ & $a^{(1)}$ & $\{\tilde M_{1a^{(1)}}$ & $\cdots$ & $\tilde M_{pa^{(1)}}\}$ & $\E\left(Y_{a^{(0)}\tilde M_{1a^{(1)}}\cdots\tilde M_{pa^{(1)}}}|L\right)$ \\\hline 
 $0$ & $0$ & $\{\tilde M_{10}$ & $\cdots$ & $\tilde M_{p0}\}$ & $\hat h(0,\tilde M_{10},\ldots,\tilde M_{p0},L)$ \\ 
 $0$ & $1$ & $\{\tilde M_{11}$ & $\cdots$ & $\tilde M_{p1}\}$ & $\hat h(0,\tilde M_{11},\ldots,\tilde M_{p1},L)$ \\ 
 $1$ & $1$ & $\{\tilde M_{11}$ & $\cdots$ & $\tilde M_{p1}\}$ & $\hat h(1,\tilde M_{11},\ldots,\tilde M_{p1},L)$ \\ 
\hline   
\end{tabular}
\end{center}
\end{table}

\item For each row of Table~\ref{table:duplicatedall_fitM1}, randomly draw the counterfactual mediator values $\{\tilde M_{1a^{(1)}} \cdots \tilde M_{pa^{(1)}}\}$ jointly from the observed treatment group $A=a^{(1)}$. 
Because treatment is randomly assigned, there are no confounders of the treatment and the (counterfactual) mediators, and assumption \eqref{eq:identify_3} is satisfied unconditionally on $L$.
Randomly select an individual whose observed treatment is $A=a^{(1)}$, then set the counterfactual mediators $\{\tilde M_{1a^{(1)}} \cdots \tilde M_{pa^{(1)}}\}$ to the selected individual's observed values $\{M_{1} \cdots M_{p}\}$.

\item Impute the expected potential outcomes as predictions $\hat h(a^{(0)},\tilde M_{1a^{(1)}},\ldots, \tilde M_{pa^{(1)}}, L)$ from the fitted outcome model in step A0.

\item Repeat steps A2 and A3 to account for the variability in the (counterfactual) mediator values, thereby obtaining the (Monte Carlo averaged) imputed potential outcomes $\E\left(Y_{a^{(0)}\{\tilde M_{1a^{(1)}}\cdots\tilde M_{pa^{(1)}}\}}|L\right)$ for each individual.

\item For each unique value of $\{a^{(0)}, a^{(1)}\}$ in Table~\ref{table:duplicatedall_fitM1}, calculate the average imputed potential outcome $\E\left(Y_{a^{(0)}\{\tilde M_{1a^{(1)}}\cdots\tilde M_{pa^{(1)}}\}}\right)$ across all individuals in the observed sample, which in doing so, averages over the empirical distribution of the covariates $L$. The estimators of the direct effect \eqref{eq:DE_define_marginal} and joint indirect effect \eqref{eq:IE_define_marginal} are obtained by plugging in the sample averages for the unknown (population) quantities.

\een

Next, we estimate the separate indirect effects via each mediator. In general, the potential outcome $Y_{a^{(0)}\tilde M_{1a^{(1)}}\cdots\tilde M_{pa^{(p)}}}$ is unobservable, even when the hypothetical treatments all equal the observed treatment (i.e., $a^{(0)}=a^{(1)}=\ldots=a^{(p)}=A$). This is because each counterfactual mediator $\tilde M_{sa^{(s)}}$, even under the observed treatment $a^{(s)}=A$, has to be drawn from its marginal (counterfactual) distribution that does not depend on any other mediators by definition. In contrast, the observed values $\{M_1 \cdots M_p\}$ are jointly distributed according to some (unknown) distribution for all the mediators. 
Estimating the indirect effects via each mediator therefore requires randomly sampling (counterfactual) mediator values from their marginal distributions, and proceeds as follows:

\ben[label=B\arabic*.]
\item Construct the duplicated data for each individual as shown in Table~\ref{table:duplicatedall_fitM2_M3}. 
Set all hypothetical treatments in the first row to $0$; i.e., $a^{(0)}=a^{(1)}=\ldots=a^{(p)}=0$.
For $s=1,\ldots,p$, set the hypothetical treatments in row $s+1$ to those in the first term of the interventional indirect effect via mediator $M_s$ as defined in \eqref{eq:IEs_define_marginal}; i.e.,
$a^{(s)}=1$ in row $s+1$, and $0$ otherwise.
The hypothetical treatment levels are chosen so that the interventional indirect effect via each mediator $M_s$ corresponds to the difference between the (transformed) estimands in rows $s+1$ and $1$.
In the last row, set $a^{(0)}=0$ and $a^{(1)}=\ldots=a^{(p)}=1$.
The interventional indirect effect via the mediators' mutual dependence \eqref{eq:IEmutual_define_marginal} is thus the difference between (i) the difference in (transformed) estimands in the penultimate and first rows of Table~\ref{table:duplicatedall_fitM1}, and (ii) the difference in (transformed) estimands in the last and first rows of Table~\ref{table:duplicatedall_fitM2_M3}.

\begin{table}
\caption{Duplicated data for each individual when estimating the separate indirect effects via each mediator. There are $t+2$ rows for a binary treatment $A$. The counterfactual mediators $\tilde M_{sa^{(s)}}$ are randomly drawn from the observed treatment group $A=a^{(s)}, s=1,\ldots,p$. The column $L$ is omitted for simplicity. \label{table:duplicatedall_fitM2_M3}}
\begin{center}
\setlength{\tabcolsep}{4pt}
\begin{tabular}{ |cccccc|ccccc|c|} 
 \hline
 $a^{(0)}$ & $a^{(1)}$ & $a^{(2)}$ & $\cdots$ & $a^{(p-1)}$ & $a^{(p)}$
 & $\tilde M_{1a^{(1)}}$ & $\tilde M_{2a^{(2)}}$ & $\cdots$ & $\tilde M_{p-1,a^{(p-1)}}$ & $\tilde M_{pa^{(p)}}$ & 
 $\E\left(Y_{a^{(0)}\tilde M_{1a^{(1)}}\cdots\tilde M_{pa^{(p)}}}|L\right)$ \\\hline 
 $0$ & $0$ & $0$ & $\cdots$ & $0$ & $0$
 & $\tilde M_{10}$ & $\tilde M_{20}$ & $\cdots$ & $\tilde M_{p-1,0}$ & $\tilde M_{p0}$
 & $\hat h(0,\tilde M_{10},\ldots,\tilde M_{p0},L)$ \\  
 $0$ & $1$ & $0$ & $\cdots$ & $0$ & $0$
 & $\tilde M_{11}$ & $\tilde M_{20}$ & $\cdots$ & $\tilde M_{p-1,0}$ & $\tilde M_{p0}$
 & $\hat h(0,\tilde M_{11},\ldots,\tilde M_{p0},L)$ \\ 
 $0$ & $0$ & $1$ & $\cdots$ & $0$ & $0$
 & $\tilde M_{10}$ & $\tilde M_{21}$ & $\cdots$ & $\tilde M_{p-1,0}$ & $\tilde M_{p0}$
 & $\hat h(0,\tilde M_{10},\ldots,\tilde M_{p0},L)$ \\ 
 $\vdots$ & $\vdots$ & $\vdots$ & $\vdots$
 & $\vdots$ & $\vdots$ & $\vdots$ & $\vdots$ & $\vdots$ & $\vdots$
 & $\vdots$ & $\vdots$ \\ 
 $0$ & $0$ & $0$ & $\cdots$ & $1$ & $0$
 & $\tilde M_{10}$ & $\tilde M_{20}$ & $\cdots$ & $\tilde M_{p-1,1}$ & $\tilde M_{p0}$ 
 & $\hat h(0,\tilde M_{10},\ldots,\tilde M_{p0},L)$ \\ 
 $0$ & $0$ & $0$ & $\cdots$ & $0$ & $1$
 & $\tilde M_{10}$ & $\tilde M_{20}$ & $\cdots$ & $\tilde M_{p-1,0}$ & $\tilde M_{p1}$ 
 & $\hat h(0,\tilde M_{10},\ldots,\tilde M_{p1},L)$ \\ 
 $0$ & $1$ & $1$ & $\cdots$ & $1$ & $1$
 & $\tilde M_{11}$ & $\tilde M_{21}$ & $\cdots$ & $\tilde M_{p-1,1}$ & $\tilde M_{p1}$ 
 & $\hat h(0,\tilde M_{11},\ldots,\tilde M_{p1},L)$ \\  
\hline   
\end{tabular}
\end{center}
\end{table}

\item In each row of Table~\ref{table:duplicatedall_fitM2_M3}, for column $s=1,\ldots,p$, randomly sample the counterfactual mediator $\tilde M_{sa^{(s)}}$, unconditionally on the other mediators, from the observed treatment group $A=a^{(s)}$. 
This can be carried out by randomly selecting an individual whose observed treatment is $A=a^{(s)}$, then setting the counterfactual mediator $\tilde M_{sa^{(s)}}$ to the selected individual's observed value $M_{s}$.
Because the randomly assigned treatment is jointly independent of all the (counterfactual) mediators when assumption \eqref{eq:identify_3} holds, it implies that the treatment is marginally independent of each (counterfactual) mediator, unconditionally on $L$.

\item Impute the expected potential outcomes as a prediction $\hat h(0,\tilde M_{1a^{(1)}},\ldots, \tilde M_{pa^{(p)}}, L)$ from the fitted outcome model in step A0.

\item Repeat steps B2 and B3 to account for the variability in the (counterfactual) mediator values, thereby obtaining the (Monte Carlo averaged) imputed potential outcomes $\E\left(Y_{a^{(0)}\tilde M_{1a^{(1)}}\cdots\tilde M_{pa^{(1)}}}|L\right)$ for each individual.

\item For each unique value of $\{a^{(0)}, a^{(1)}, \ldots, a^{(p)}\}$ in Table~\ref{table:duplicatedall_fitM2_M3}, calculate the average imputed potential outcome $\E\left(Y_{a^{(0)}\tilde M_{1a^{(1)}}\cdots\tilde M_{pa^{(p)}}}\right)$ across all individuals in the observed sample. The estimators of the interventional indirect effect via each mediator $M_s$ \eqref{eq:IEs_define_marginal} are obtained by plugging in the sample averages for the unknown (population) quantities.
The estimator of the indirect effect via the mediators' mutual dependence \eqref{eq:IEmutual_define_marginal} is similarly obtained using the sample averages in the second and first rows of Table~\ref{table:duplicatedall_fitM1}, and the last and first rows of Table~\ref{table:duplicatedall_fitM2_M3}.

\een

Unbiased estimation of the interventional effects therefore depends on correctly specifying an outcome model conditional on treatment, mediators, and covariates that is unbiased for the marginal structural mean model; i.e., 
$\hat h(a,m_1, \ldots, m_p, L)$ converges in probability to $\E\left(Y_{a m_1\cdots m_p}|L\right)$.
The consistency of the estimator under the identifying assumptions \eqref{eq:identify_1}--\eqref{eq:identify_3} is shown in Web Appendix~\ref{sect:consistency}.
Standard errors can be estimated using a non-parametric percentile bootstrap procedure \citep{efron1994introduction} that randomly resamples observations with replacement, then repeating steps A0 -- A5 and B1 -- B5 for each bootstrap sample. 

\subsection{High-dimensional mediators \label{sect:highdimensional_doubleselection}}

In this section, we consider high-dimensional mediation settings with fewer observations than mediators (and covariates).
Separate mediation analyses can lead to biased estimates of indirect effects because merely omitted mediators can act as unobserved (post-treatment) confounders of the mediator(s) in question and the outcome, thus violating assumption \eqref{eq:identify_2}. 
Recall the example in Figure~\ref{fig-hdexample}, where $M_1$ and $M_2$ are confounders of the $M_3-Y$ relation. A single mediation analysis for $M_3$ that ignores either $M_1$ or $M_2$, or both, will yield biased estimates of the indirect effect via $M_3$. 
In this section, we propose a strategy to estimate the separate indirect effects via each mediator, by focusing on each distinct mediator $M_s, s=1,\ldots,p$, in turn. For the mediator $M_s$ in question, possible confounding of the mediator-outcome relation (by other mediators or possibly high-dimensional covariates) is carefully adjusted for to ensure unbiased estimation of the interventional indirect effect.

\subsubsection{Outcome model} 
First, fit a penalized regression model for the outcome to the observed data. 
Here and throughout, we will consider an elastic net penalty \citep{zou2005regularization}, which is a compromise between the ridge regression penalty and the lasso penalty.
The elastic net penalty is especially useful when there are many correlated predictor variables, and has been applied to different gene expression datasets with highly correlated genes \citep{glmnetR2010}.
Because evaluating the indirect effect via $M_s$ requires jointly assessing the $M_s-Y$ and $A-M_s$ relations, (prematurely) shrinking the coefficient of $M_s$ to zero can lead to biased inference. 
The coefficients for $A$ and $M_s$ should therefore be unpenalized, e.g., by setting their penalty factors to zero, to retain both $A$ and $M_s$ in the outcome model. 
Denote the subset of the remaining mediators and covariates with non-zero (penalized) coefficients by ${\cal M}_s(Y)$, where the subscript emphasizes the focus on mediator $M_s$. 

\subsubsection{Mediator model} 
Next, to increase the chances of detecting confounders of the $M_s-Y$ relation, fit a penalized regression model (using an elastic net penalty) for $M_s$ conditional on all other mediators, treatment, and covariates to the observed data.
Denote the subset of the predictors with non-zero (penalized) coefficients by ${\cal M}_s(M)$.

\subsubsection{Indirect effect via each distinct mediator}
Denote the union of the selected predictors (with non-zero coefficients) in either the penalized outcome model or the penalized mediator model by ${\cal M}_s = {\cal M}_s(Y) \bigcup {\cal M}_s(M)$.
The selected mediators and covariates in ${\cal M}_s$ are thus possible confounders of the $M_s - Y$ relation, and can be included in a new outcome model. Denote this outcome model by $h(A,M_s, {\cal M}_s)$, where the dependence on only the selected mediators (and covariates) in ${\cal M}_s$ is implied. 
To estimate the indirect effect via $M_s$, carry out steps B1 -- B5 using only rows $s+1$ and $1$ of Table~\ref{table:duplicatedall_fitM2_M3}. Predictions from the (unpenalized) outcome model $h(A,M_s, {\cal M}_s)$ fitted to the observed data are used to impute potential outcomes in step B4. 

\subsubsection{Remarks}
Separately selecting predictors of the outcome in ${\cal M}_s(Y)$ and predictors of the mediator in ${\cal M}_s(M)$ can increase the chances of selecting (observed) confounders of the $M_s-Y$ relation. Continuing the example in Figure~\ref{fig-hdexample}, suppose that $M_2$ is omitted from ${\cal M}_3(Y)$ due to sampling variability, but selected in ${\cal M}_3(M)$. Then $M_2 \in {\cal M}_3$ would still be adjusted for in the outcome model $h(A, M_3, {\cal M}_3)$.
The use of separate outcome and mediator models is inspired by {\em double selection} principles \citep{belloni2014inference}, where in settings without mediators, the (partial) associations between the outcome and the covariates, and between the treatment and the covariates, are evaluated. 
We acknowledge that specifying mediator models may appear contradictory to the key feature of our proposal that avoids modeling the relations between the mediators. 
But we emphasize that the (penalized) regression model for each mediator in question is used merely to identify possible confounders of the mediator-outcome relation, by essentially viewing the other mediators as possible confounders. These models do not assume (or empirically imply) any causal structure among the mediators.

\subsection{Observational studies with non-randomly assigned treatments \label{sect:observational}}

When treatment is not randomly assigned, the counterfactual mediator values used to construct the duplicated data in the estimation procedure cannot be sampled by merely selecting mediator values at random within each observed treatment group.
The observed baseline confounders $L$ of the treatment-mediator(s) and treatment-outcome relations have to be adjusted for toward ensuring that the identifying assumptions \eqref{eq:identify_1} and \eqref{eq:identify_3} hold.
For example, the counterfactual density for mediator $M_s$ when setting $A$ to $a$ is:
\begin{align*}
f(M_{sa})
&=\int f(M_{s}|A=a,L=l) f(L=l)f(L=l|A=a)^{-1} \, dF_{L|A=a}(l) \\
&=\Pr(A=a)\int f(M_{s}|A=a,L=l) \Pr(A=a|L=l)^{-1} \, dF_{L|A=a}(l), \quad a=0,1.
\end{align*}
The above result motivates a modified estimation procedure for non-randomly assigned treatments, by sampling the (observed) mediator values with probability proportional to the inverse of the conditional probability of receiving the observed treatment given the confounders.
For example, suppose that $a^{(s)}=0$. Consider an individual in the treatment group $A=0$ with covariate values $L=l$ and observed mediator value $M_{s}$. The sampling probability of this particular individual's value of $M_{s}$ is $\Pr(A=0|L=l)^{-1} / \{ \sum_{i} \Pr(A=0|L=l^i)^{-1} \indicator{A^i=0}\}$, where the sum in the denominator is over all individuals $i$ (as indexed by the $i$ superscripts without brackets) with covariate values $L=l^i$ in the (same) treatment group $A^i=0$. 
The (counterfactual) mediator value(s) can be readily sampled by randomly selecting an individual from the treatment group $A=a^{(s)}$ according to the aforementioned sampling probabilities, then setting the counterfactual mediator to the selected individual's observed mediator value(s).
In practice, the probabilities of the observed treatments $\Pr(A=a|L), a=0,1$, may be estimated using predictions from a (saturated) logistic regression model fitted to the observed data, with the treatment as the dependent variable, and main effects (and when feasible, higher order and interaction effects) for the covariates as the predictors.

\section{Simulation studies \label{sect:simstudies}}

Two simulation studies were conducted to empirically assess the operating characteristics of the proposed estimation strategy in finite samples across different settings. Details of the procedures and results of these simulation studies are deferred to Web Appendix~\ref{sect:simstudies_details}. To provide an overview, in study 1, we considered a setting with two (continuous) mediators where one was affected by the other.
The estimators of the interventional effects proposed in this article were compared with existing estimators of natural effects \citep{steen2017flexible} that required stricter identification assumptions. 
Settings where the causal ordering of the mediators was either correctly or incorrectly assumed, and unobserved confounding of the mediators was either present or absent, were considered.
In study 2, a setting with high-dimensional (non-continuous) mediators was considered for assessing the indirect effect estimators proposed in Section~\ref{sect:highdimensional_doubleselection}.
The outcome was binary in both studies.

The results of the simulation studies showed that estimators of the existing natural indirect effects were unbiased only when the identifying assumptions were met; i.e., the mediators' causal ordering was correctly assumed, and there was no unobserved confounding of the mediators. As expected, when at least one of the assumptions was violated, estimates were biased even at large sample sizes. 
In contrast, estimators of the interventional effects proposed in this article were unbiased under all the considered settings.
The proposed estimation strategy for high-dimensional mediators was able to detect the mediators through which the indirect effects were transmitted with high probability empirically. The estimated indirect effects were empirically biased at smaller sample sizes, but the biases tended to zero as the sample sizes increased.

\section{Illustration with an example dataset \label{sect:application}}

We illustrated the proposed estimation strategy using publicly available data from a previous high-dimensional mediation analysis by \citet{huang2016hypothesis}.
\citeauthor{huang2016hypothesis} assessed whether the causal effect of the microRNA miR-223 expression (the treatment of interest) on mortality within three months due to glioblastoma multiforme (GBM), a malignant brain tumor, was mediated by expression values of different genes in the tumor genome.
Expression values for 1220 genes (with no missing data) from 490 patients suffering from GBM were included in the online supplemental materials of \citet{huang2016hypothesis}.
Interventional effects are suitable for such high-dimensional mediator settings because the mediators' internal causal structure, such as gene regulation networks or protein signaling networks, are often unknown in practice. 
For the purposes of illustrating the proposed estimation strategy, we made the following simplifying assumptions. 
We dichotomized the miR-223 expression at the empirical median so that $A=1$ if the expression was above the median, or 0 otherwise.
We used a dichotomous indicator of whether death from GBM had occurred in the first three months as the outcome ($Y$). 
Among the 490 patients, 44 died within the first three months ($Y=1$).
We assumed that the patients lost to follow-up prior to three months (four in the $A=0$ group, and five in the $A=1$ group) were alive ($Y=0$). 
We assumed that the baseline demographic variables in the publicly available data (age at diagnosis, gender, and ethnicity), which we jointly denoted by $L$, were sufficient for the identifying assumptions \eqref{eq:identify_1}--\eqref{eq:identify_3} to hold. In practice, a richer set of (baseline) covariates should be adjusted for in substantive analyses to reduce the possibility of biases due to unobserved confounding. 

\begin{figure}
	\centering
	\includegraphics[width=\linewidth]{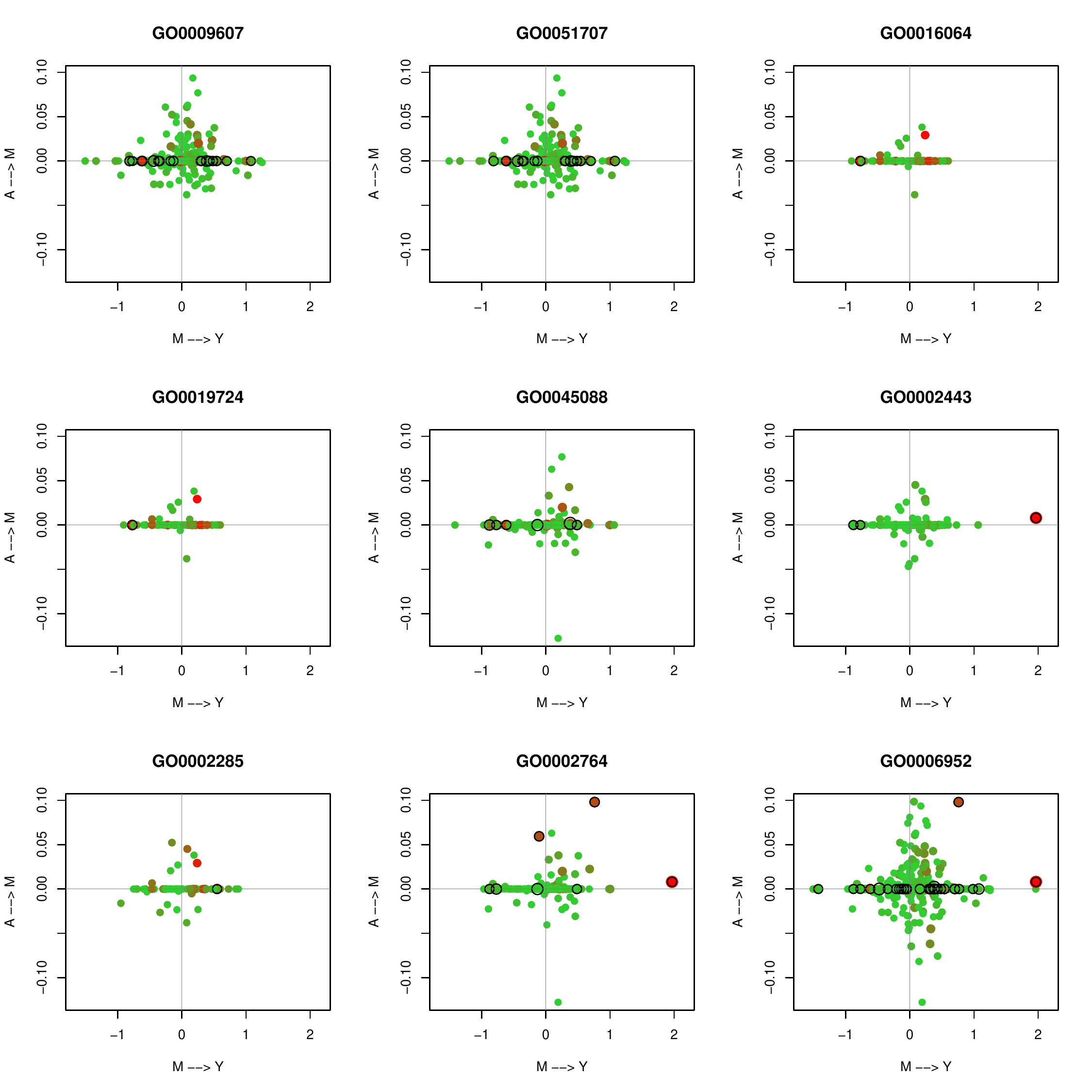}
	\caption{Scatterplots of the (penalized) regression coefficients in the linear mediator and logistic outcome models for the GBM data set. Each point corresponds to a distinct mediator, with the coefficient of treatment on the mediator in question on the vertical axis (``$A \rightarrow M$''), and the coefficient of the mediator on the outcome on the horizontal axis (``$M \rightarrow Y$''). The size and color of the points are proportional to the (absolute) magnitudes of the indirect effects, with larger points and darker (red) hues indicating indirect effect estimates further from zero. Mediators whose 95\% CIs excluded zero are circled in black.
	Each panel corresponds to one of the nine sets of mediators described in \citeauthor{huang2016hypothesis}.
	\label{plot-dataex-iesets}
	}
	\end{figure}

The proposed estimation strategy for high-dimensional mediators described in Section~\ref{sect:highdimensional_doubleselection} was carried out. 
For the purposes of selecting confounders, linear models were assumed for all (continuous) mediators, and a logistic regression model was assumed for the (binary) outcome.
The penalized regression models using elastic net penalties were fitted using the \texttt{glmnet} package \citep{glmnetR2010} in \texttt{R}.
The value of the tuning parameter that controlled the overall strength of the penalty in each model was selected using 10-fold cross-validation (the default setting) on either the mean squared error (for the continuous mediators) or the misclassification error (for the binary outcome).
We again emphasize that the mediator models were used merely to select possible confounders of the mediator-outcome relations that required adjustment (in the subsequent unpenalized outcome model) to avoid biases due to unobserved confounding.
Estimating each indirect effect only requires an outcome model fitted to the observed data, and does not require assuming any model for the mediators.
Because treatment was not randomly assigned, counterfactual mediator values were sampled using the proposed weights in Section~\ref{sect:observational} under a saturated logistic regression model for treatment conditional on the baseline covariates.
For each mediator, we plotted the (penalized) estimated regression coefficients of treatment in the linear mediator model, and of the mediator in the logistic outcome model, in Figure~\ref{plot-dataex-iesets}. 
Each panel corresponded to one of the nine (overlapping) sets of mediators defined in Table 2 of \citet{huang2016hypothesis}. We refer readers to \citeauthor{huang2016hypothesis} for details of the biological functions of each gene set.

Non-parametric bias-corrected and accelerated 95\% confidence intervals (CIs) were constructed, by applying the \texttt{bca} function \citep{coxedR} to $500$ bootstrap samples that randomly resampled observations with replacement and repeated the estimation procedure for each sample.
Genes whose 95\% CIs excluded zero are displayed in Table~\ref{table:gbm_results}. 
The interventional indirect effect via an individual gene can be interpreted using the following example. The odds of death from GBM within three months due to shifting the marginal distribution of the PIK3CD gene expression from those with observed miR-223 expressions above the median to those below the median, while holding the distributions of all other gene expressions fixed among those with miR-223 expressions below the median, was estimated to increase by $\exp(0.62) \approx 1.85$ times $(95\%$ CI $= (1.20, 2.61))$.
To compare our results with those in \citeauthor{huang2016hypothesis}, we indicated in Table~\ref{table:gbm_results} which gene set(s) each distinct gene belonged to. \citeauthor{huang2016hypothesis} found that all nine gene sets significantly mediated the causal effect of miR-223 expression on GBM mortality within three months (with p-values less than 0.05). We (similarly) found that each gene set contained at least one gene with a (significantly) non-zero interventional indirect effect.
Because several simplifying assumptions were made for the sole purpose of illustrating the proposed estimation strategy, substantive conclusions about the biological significance of the statistically significant mediators are beyond the scope of this article.

\begin{table}[ht]
\caption{Estimates (``Est.'') on the log-odds scale, and non-parametric bias-corrected and accelerated bootstrap 95\% confidence intervals (``CIs''), for the interventional indirect effects via each distinct gene in the GBM data set. 
Only genes whose 95\% CIs excluded zero are displayed. The genes are ordered by increasing lower bound of their 95\% CI.
Each gene either belongs  (``$\checkmark$'') or does not belong (``--'') to a gene set described in \citeauthor{huang2016hypothesis}.
The gene sets in the column headings are labelled as (i) GO0009607, (ii) GO0051707, (iii) GO0016064, (iv) GO0019724, (v) GO0045088, (vi) GO0002443, (vii) GO0002285, (viii) GO0002764, and (ix) GO0006952.
All results are rounded to two decimal places. \label{table:gbm_results}}
\centering
\begin{tabular}{lrrrrrrrrrrrr}
  \hline
 Gene & Est. & \multicolumn{2}{c}{95\% CI} & (i) & (ii) & (iii) & (iv) & (v) & (vi) & (vii) & (viii) & (ix) \\ 
  \hline

FYB & -0.44 & -1.16 & -0.28 & -- & -- & -- & -- & -- & -- & -- & $\checkmark$ & -- \\ 
  TMEM173 & -0.37 & -0.99 & -0.07 & $\checkmark$ & $\checkmark$ & -- & -- & $\checkmark$ & -- & -- & -- & $\checkmark$ \\ 
  PRDM1 & -0.14 & -0.82 & -0.15 & $\checkmark$ & $\checkmark$ & -- & -- & -- & -- & -- & -- & -- \\ 
  SAA1 & -0.09 & -0.71 & -0.08 & -- & -- & -- & -- & -- & -- & -- & -- & $\checkmark$ \\ 
  SAA2 & -0.10 & -0.69 & -0.09 & -- & -- & -- & -- & -- & -- & -- & -- & $\checkmark$ \\ 
  FCN2 & -0.33 & -0.66 & -0.16 & -- & -- & -- & -- & -- & -- & -- & -- & $\checkmark$ \\ 
  IRF5 & -0.24 & -0.62 & -0.09 & $\checkmark$ & $\checkmark$ & -- & -- & -- & -- & -- & -- & $\checkmark$ \\ 
  ARRB2 & -0.22 & -0.58 & -0.08 & -- & -- & -- & -- & $\checkmark$ & $\checkmark$ & -- & $\checkmark$ & $\checkmark$ \\ 
  SPON2 & -0.11 & -0.55 & -0.03 & -- & -- & -- & -- & -- & -- & -- & -- & $\checkmark$ \\ 
  NUPR1 & -0.04 & -0.54 & -0.01 & -- & -- & -- & -- & -- & -- & -- & -- & $\checkmark$ \\ 
  SAA4 & -0.35 & -0.48 & -0.05 & -- & -- & -- & -- & -- & -- & -- & -- & $\checkmark$ \\ 
  CD226 & -0.22 & -0.48 & -0.10 & $\checkmark$ & -- & $\checkmark$ & $\checkmark$ & $\checkmark$ & $\checkmark$ & -- & $\checkmark$ & $\checkmark$ \\ 
  NCAM1 & -0.18 & -0.48 & -0.03 & -- & -- & -- & -- & -- & -- & -- & -- & $\checkmark$ \\ 
  UACA & -0.12 & -0.43 & -0.02 & -- & -- & -- & -- & -- & -- & -- & -- & $\checkmark$ \\ 
  CYP11A1 & -0.10 & -0.39 & -0.01 & $\checkmark$ & $\checkmark$ & -- & -- & -- & -- & -- & -- & -- \\ 
  FOS & -0.06 & -0.38 & -0.00 & $\checkmark$ & $\checkmark$ & -- & -- & $\checkmark$ & -- & -- & $\checkmark$ & $\checkmark$ \\ 
  GUCY1A3 & -0.05 & -0.31 & -0.00 & $\checkmark$ & $\checkmark$ & -- & -- & -- & -- & -- & -- & -- \\ 
  LGALS8 & 0.25 & 0.00 & 0.57 & -- & -- & -- & -- & -- & -- & $\checkmark$ & -- & -- \\ 
  TXN & 0.10 & 0.00 & 0.34 & -- & -- & -- & -- & -- & -- & -- & -- & $\checkmark$ \\ 
  CCDC130 & 0.13 & 0.00 & 0.36 & $\checkmark$ & $\checkmark$ & -- & -- & -- & -- & -- & -- & -- \\ 
  DEFA5 & 0.11 & 0.00 & 0.44 & $\checkmark$ & $\checkmark$ & -- & -- & -- & -- & -- & -- & $\checkmark$ \\ 
  APOBEC3G & 0.18 & 0.01 & 0.63 & $\checkmark$ & $\checkmark$ & -- & -- & -- & -- & -- & -- & $\checkmark$ \\ 
  IL17RA & 0.13 & 0.01 & 0.59 & -- & -- & -- & -- & -- & -- & -- & -- & $\checkmark$ \\ 
  CCL4 & 0.15 & 0.03 & 0.47 & $\checkmark$ & $\checkmark$ & -- & -- & -- & -- & -- & -- & $\checkmark$ \\ 
  TCF3 & 0.14 & 0.03 & 0.51 & $\checkmark$ & $\checkmark$ & -- & -- & -- & -- & -- & -- & -- \\ 
  LSP1 & 0.24 & 0.04 & 0.67 & -- & -- & -- & -- & -- & -- & -- & -- & $\checkmark$ \\ 
  CFH & 0.11 & 0.04 & 0.50 & -- & -- & -- & -- & -- & -- & -- & -- & $\checkmark$ \\ 
  TAPBP & 0.09 & 0.04 & 0.60 & -- & -- & -- & -- & -- & -- & -- & -- & $\checkmark$ \\ 
  CTSB & 0.25 & 0.05 & 0.91 & -- & -- & -- & -- & -- & -- & -- & -- & $\checkmark$ \\ 
  FZD5 & 0.12 & 0.05 & 0.40 & $\checkmark$ & $\checkmark$ & -- & -- & -- & -- & -- & -- & -- \\ 
  TSPO & 0.19 & 0.05 & 0.63 & $\checkmark$ & $\checkmark$ & -- & -- & -- & -- & -- & -- & -- \\ 
  SPHK1 & 0.20 & 0.06 & 0.56 & -- & -- & -- & -- & -- & -- & -- & -- & $\checkmark$ \\ 
  GBP3 & 0.21 & 0.12 & 0.52 & $\checkmark$ & $\checkmark$ & -- & -- & -- & -- & -- & -- & $\checkmark$ \\ 
  IRF1 & 0.18 & 0.13 & 0.61 & $\checkmark$ & $\checkmark$ & -- & -- & $\checkmark$ & -- & -- & $\checkmark$ & $\checkmark$ \\ 
  A2M & 0.29 & 0.14 & 0.75 & -- & -- & -- & -- & $\checkmark$ & -- & -- & -- & $\checkmark$ \\ 
  PIK3CD & 0.62 & 0.18 & 0.96 & -- & -- & -- & -- & -- & $\checkmark$ & -- & $\checkmark$ & $\checkmark$ \\ 
  CLEC5A & 0.39 & 0.25 & 0.82 & $\checkmark$ & $\checkmark$ & -- & -- & -- & -- & -- & -- & $\checkmark$ \\ 
  ITK & 0.43 & 0.25 & 0.85 & -- & -- & -- & -- & -- & -- & -- & $\checkmark$ & $\checkmark$ \\ 
   \hline   
\end{tabular}
\end{table}

\section{Comparison \label{sect:compare}}

In this section, we argue that in common settings, the interventional effects proposed in this article are equal to existing definitions by \citet{vansteelandt2017interventional}.
In particular, the existing definitions rely on (counterfactual) mediator distributions that depend on treatment and all baseline covariates, including confounders of the mediator(s)-outcome relation(s).
Suppose that all mediators are continuous, so that linear mean models for the mediators (conditional on $L$) may be assumed.
Suppose that the outcome is dichotomous, and sufficiently rare so that the logit link in a (marginal structural) mean model for the outcome can be approximated by the log link. When (i) the outcome model has main effects only, and (ii) the effect of treatment on the mediators is not moderated by the confounders (i.e., there are no interaction terms involving treatment in the mediator mean models), the proposed interventional effects are equal to existing definitions. 
Detailed results are provided in Web Appendix~\ref{sect:comparison}.

\section{Conclusion \label{sect:discussion}}

In this article, we have developed interventional effects for high-dimensional mediators that permit a novel estimation strategy allowing for any (marginal structural) mean model for the outcome, and requiring no models for the mediators. 
There are several avenues of possible future related research.
The proposed estimation strategy for high-dimensional mediators is designed to deliver reasonable approximations of the indirect effects through selecting confounders (including other mediators) using the double selection principle. Establishing theoretical properties to ensure valid inference of the proposed Monte Carlo-based estimators, especially following mediator selection, is an area of future work. When selecting a subset of the mediators for further investigation, a threshold that either is pre-determined, such as in \citet{huang2016hypothesis}, or controls the familywise error rate under multiple testing scenarios, such as in \citet{derkach2020group}, can be used.
When further interest is in estimating the direct, joint, or mutual indirect effects, a separate penalized regression model for the outcome (conditional on all the mediators, covariates, and treatment) must be fitted to the observed data for predicting the potential outcomes in Table~\ref{table:duplicatedall_fitM1}, and in the first and last rows of Table~\ref{table:duplicatedall_fitM2_M3}, of the estimation strategy in Section~\ref{sect:lowdimensional_random}. 
The estimated interventional indirect effects via each distinct mediator can be used to determine the penalties; e.g., the indirect effects can be used in place of the product of coefficients in the penalties for the pathway lasso \citep{zhao2016pathway}.
High-dimensional mediation methods that employ principal components analysis, such as \citet{huang2016hypothesis} and \citet{zhao2020sparse}, can be used to select mediators with high loadings in the principal components. Substantive comparisons of the indirect effects through specific selected (sets of) mediators using such approaches, with the interventional indirect effects developed in this article, may be made in future work.
Unbiased estimation of the interventional effects defined in this article depends on correctly specifying a mean model for the outcome. \citet{vansteelandt2012imputation} recommend using sufficiently rich outcome models that e.g., include higher-order or interaction terms between treatment and mediators, or between mediators. 
Assessing robustness to misspecification of the outcome model, e.g., due to omitting such interaction terms, is a direction for future research.
Other more general non-parametric prediction methods for the potential outcomes that leverage data-adaptive techniques \citep{diaz2019non, benkeser2020nonparametric} can also be considered. 
Extending such machine learning methods to high-dimensional settings requires further work to avoid the need for inverse weighting by the joint mediator density.
			
\backmatter

\vspace*{-.1in}

\section*{Acknowledgements}
The authors would like to thank the Editor, Associate Editor, and two reviewers for their comments on prior versions of this manuscript. 
This research was supported by the Research Foundation -- Flanders (FWO) under Grant G019317N. 
Computational resources and services were provided by the VSC (Flemish Supercomputer Center), funded by the FWO and the Flemish Government -- department EWI.
The content is solely the responsibility of the authors and does not 
represent the official views of the authors' institutions or FWO.

\vspace*{-.2in}

\bibliographystyle{biom}
\bibliography{interventional-highdim}


\vspace*{-.3in}

\section*{Supporting Information}

Web Appendices \ref{sect:consistency} to \ref{sect:comparison} referenced in Sections \ref{sect:notation_definition} to \ref{sect:compare} are available with this paper at the Biometrics website on Wiley Online Library.
The \texttt{R} code used to implement the proposed estimation procedure and to carry out the simulation studies in Section~\ref{sect:simstudies}, and the illustration in Section~\ref{sect:application}, are available at:
\url{
https://github.com/wwloh/interventional-hdmed
}


\end{document}